\documentclass[fleqn,10pt]{wlscirep}
\usepackage[utf8]{inputenc}
\usepackage[T1]{fontenc}
\usepackage{bm}
\usepackage{graphicx}
\usepackage{xcolor}
\usepackage{xfrac}
\usepackage{lineno}
\usepackage{ulem}

\newcommand{\CMAO}{CeMgAl$_{11}$O$_{19}$}
\newcommand{\PMAO}{PrMg(Zn)Al$_{11}$O$_{19}$}
\newcommand{\RMAO}{REMg(Zn)Al$_{11}$O$_{19}$}

\DeclareMathAlphabet\mathbfcal{OMS}{cmsy}{b}{n}

\title{Spin Excitation Continuum in the Exactly Solvable Triangular-Lattice Spin Liquid \CMAO}

\author[1,17]{Bin Gao}
\author[1,2,17]{Tong Chen}
\author[3,17]{Chunxiao Liu}
\author[1]{Mason L. Klemm}
\author[4]{Shu Zhang}
\author[5,*]{Zhen Ma}
\author[6,7,++]{Xianghan Xu}
\author[8]{Choongjae Won}
\author[9]{Gregory T. McCandless}
\author[10]{Naoki Murai}
\author[10]{Seiko Ohira-Kawamura}
\author[11]{Stephen J. Moxim}
\author[11]{Jason T. Ryan}
\author[12]{Xiaozhou Huang}
\author[13]{Xiaoping Wang}
\author[9]{Julia Y. Chan}
\author[6]{Sang-Wook Cheong}
\author[2]{Oleg Tchernyshyov}
\author[14*]{Leon Balents}
\author[1,15,16*]{Pengcheng Dai}

\affil[1]{Department of Physics and Astronomy, Rice University, Houston, Texas, USA.}
\affil[2]{Institute for Quantum Matter and Department of Physics and Astronomy, Johns Hopkins University, Baltimore, Maryland, USA.}
\affil[3]{Department of Physics, University of California, Berkeley, California, USA.}
\affil[4]{Max-Planck-Institut fur Physik komplexer Systeme, Dresden, Germany.}
\affil[5]{Hubei Key Laboratory of Photoelectric Materials and Devices, School of Materials Science and Engineering, Hubei Normal University, Huangshi, China.}
\affil[6]{Rutgers Center for Emergent Materials and Department of Physics \& Astronomy, Rutgers University, Piscataway, New Jersey, USA}
\affil[7]{Department  of  Chemistry,  Princeton  University,  Princeton,  New Jersey, USA.}
\affil[8]{Laboratory for Pohang Emergent Materials and Max Planck POSTECH Center for Complex Phase Materials, Pohang University of Science and Technology, Pohang, Korea.}
\affil[9]{Department of Chemistry and Biochemistry, Baylor University, Waco, Texas, USA.}
\affil[10]{J-PARC Center, Japan Atomic Energy Agency, Tokai, Ibaraki, Japan.}
\affil[11]{Alternative Computing Group, National Institute of Standards of Technology, Gaithersburg, Maryland, USA.}
\affil[12]{Chemical Sciences and Engineering Division, Argonne National Laboratory, Lemont, Illinois, USA.}
\affil[13]{Neutron Scattering Division, Oak Ridge National Laboratory, Oak Ridge, Tennessee, USA.}
\affil[14]{Kavli Institute for Theoretical Physics, University of California, Santa Barbara, California, USA.}
\affil[15]{Smalley-Curl Institute, Rice University, Houston, Texas, USA.}
\affil[16]{Rice Center for Quantum Materials, Rice University, Houston, Texas, USA.}
\affil[17]{These authors contributed equally: Bin Gao, Tong Chen, Chunxiao Liu.}
\affil[*]{e-mail: zma@hbnu.edu.cn, balents@spinsandelectrons.com, pdai@rice.edu}
\affil[++]{Current Address: School of Physics and Astronomy, University of Minnesota, Minneapolis, Minnesota, 55455, USA}

\begin{abstract}
In magnetically ordered insulators, elementary quasiparticles manifest as spin waves - collective motions of localized magnetic moments propagating through the lattice - observed via inelastic neutron scattering. In effective spin-$\frac{1}{2}$ systems where geometric frustrations suppress static magnetic order, spin excitation continua can emerge, either from degenerate classical spin ground states or from entangled quantum spins characterized by emergent gauge fields and deconfined fractionalized excitations. Comparing the spin Hamiltonian with theoretical models can unveil the microscopic origins of these zero-field spin excitation continua. Here, we use neutron scattering to study spin excitations of the two-dimensional (2D) triangular-lattice effective spin-$\frac{1}{2}$ antiferromagnet \CMAO{}. Analyzing the spin waves in the field-polarized ferromagnetic state, we find that the spin Hamiltonian is close to an exactly solvable 2D triangular-lattice XXZ model, where degenerate 120$^\circ$ ordered ground states -- umbrella states -- develop in the zero temperature limit. We then find that the observed zero-field spin excitation continuum matches the calculated ensemble of spin waves from the umbrella state manifold, and thus conclude that \CMAO{} is the first example of an exactly solvable spin liquid on a triangular lattice where the spin excitation continuum arises from the ground state degeneracy.
\end{abstract}
\begin{document}

\flushbottom
\maketitle

\thispagestyle{empty}

\section*{Introduction}
\begin{figure}[ht]
\centering
\includegraphics[width=1.0\textwidth]{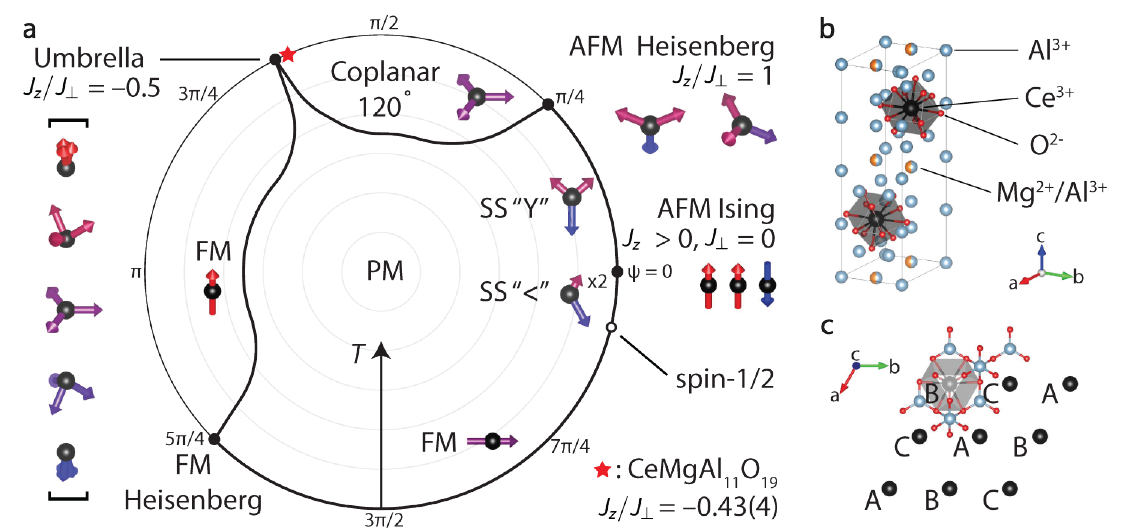}
\caption{
\textbf{Phase Diagram and Crystal Structure of \CMAO{}.}
\textbf{a}, Phase diagram of the XXZ model described by Eq. \ref{Eq:Hxxz} reproduced from ref~\cite{momoi1992ground, yamamoto2014quantum, sellmann2015phase}.
The exchange interactions are parameterized as $J_z = \cos \psi$ and $J_\perp = \sin \psi$.
Ferromagnetic, paramagnetic, and supersolid phases are labeled as ``FM'', ``PM'', and ``SS'', respectively.
The thick solid line indicating transition temperatures is decorated with phase boundaries including three spin liquids at $\psi=0$, $\frac{1}{4}\pi$, and $\psi_\textup{U}$.
The phase boundary between the coplanar FM and supersolid ``$<$'' phases (open circle) are different for spin-$\frac{1}{2}$ and classical spins, and the ground state on the boundary remains elusive. 
\CMAO{} is labeled by the red star. \textbf{b}, Crystal structure of \CMAO{},  where the CeO$_{12}$ polyhedra are highlighted. \textbf{c}, Top view of the Ce$^{3+}$ ions triangular-lattice plane.}
\label{fig1}
\end{figure}

In 1973, Anderson proposed that a spin liquid, where interacting quantum spins ($S=\frac{1}{2}$) do not exhibit long-range magnetic order even at $T=0$, is the ground state of the nearest-neighbor (NN) spin-$\frac{1}{2}$ two-dimensional (2D) triangular lattice Heisenberg antiferromagnet \cite{ANDERSON1973153}. A quantum spin liquid (QSL) is characterized by emergent gauge fields and deconfined fractionalized excitations (e.g., spinons) \cite{balents2010spin, zhou2017quantum, broholm2020quantum, savary2016quantum, mila2000quantum}. This exotic state of matter is important for our understanding of high-transition temperature superconductivity \cite{doi:10.1126/science.235.4793.1196,RevModPhys.78.17} and has potential applications in quantum computation \cite{KITAEV20032,KITAEV20062}. 
Despite decades of theoretical and experimental efforts, experimental confirmation of a QSL in real materials remains elusive \cite{balents2010spin,zhou2017quantum,broholm2020quantum}. 
Theoretically, it is now established that the ground state of an isotropic NN Heisenberg antiferromagnet on a 2D triangular lattice is not a QSL, but exhibits coplanar 120$^\circ$ magnetic order at $T=0$ (Fig.~\ref{fig1}a)  \cite{bernu1994exact, capriotti1999long, zheng2006excitation, white2007neel}.

When spin-orbit coupling of the magnetic ions becomes significant, as in rare-earth triangular-lattice antiferromagnets with an effective $S=\frac{1}{2}$ moment \cite{PhysRevLett.115.167203,WOS:000391190500046,WOS:000394070700014,Liu_2018,WOS:000488590700022,PhysRevX.11.021044,WOS:001099033900001}, the in-plane ($XY$) and out-of-plane (Ising) magnetic exchange interactions could be highly anisotropic \cite{Dorey_2007,PhysRevB.94.035107,10.21468/SciPostPhys.4.1.003}. 
This anisotropy is captured by the XXZ model Hamiltonian, which relates the in-plane ($J_\perp$) and out-of-plane ($J_z$) exchange interactions as follows:
\begin{equation}
{\cal H} = \sum_{<i,j>} [J_z  S_i^z S_j^z + J_\perp (S_i^x S_j^x + S_i^y S_j^y)],
\label{Eq:Hxxz}
\end{equation}
where the sum is over adjacent sites, and $S_k^z$ is a spin-$S$ operator.

In the limit of extreme Ising anisotropy with antiferromagnetic (AFM) interactions, three spins on each site of a triangle which point either `up' or `down' cannot simultaneously satisfy all three bonds due to geometric frustration. 
This corresponds to the classical Ising model on a triangular lattice -- $\psi=0$ in the phase diagram of the XXZ model -- shown in Figure~\ref{fig1}a, where $J_z = \cos \psi$ and $J_\perp = \sin \psi$ ($\psi$ is the angle characterizing the in-plane and out-of-plane anisotropy).
In 1950, Wannier \cite{wannier1950antiferromagnetism} and Houtappel \cite{houtappel1950order} showed that the classical AFM Ising model exhibits critical order with extensive degeneracy at $T=0$ and retains finite entropy of $0.3231R$, where $R$ is the universal gas constant, due to the degenerate spin configurations of the classical ground state \cite{PhysRevB.63.224401}. 
At low but finite temperatures, spins fluctuate thermally in a correlated manner, confined to the ground states. Analogous to an ordinary liquid, the spins in the classical Ising model on a triangular lattice form a ``spin liquid'', distinct from a QSL where emergent gauge fields and deconfined fractionalized excitations are expected \cite{balents2010spin}.

In 1986, Miyashita studied the classical XXZ model and identified the phase boundary between ferromagnetic (FM) and coplanar 120$^\circ$ phases at $\psi_{\rm U}=\pi-\arctan2 \approx 0.648\pi$, i.e. $J_z=-0.5J_\perp$, $J_\perp>0$, where the system exhibits a three-sublattice umbrella order (Fig.~\ref{fig1}a) \cite{miyashita1986ground}. 
The quantum spin-$\frac{1}{2}$ XXZ model was later explored by Momoi and Suzuki in 1992 \cite{momoi1992ground}. 
Remarkably, the spin-$\frac{1}{2}$ model at $\psi_{\rm U}$ is exactly solvable and the ground state manifold has a degeneracy coinciding with that of the classical model, featuring quantum versions of the umbrella states. It is further shown that neither FM nor coplanar 120$^\circ$ order develops at finite temperatures, resulting in a spin liquid due to thermal fluctuations amongst the degenerate umbrella states\cite{PhysRevB.84.214418,PhysRevB.96.014431,PhysRevB.93.224402,Starykh_2015,PhysRevLett.120.207203}. 
Despite the theoretical clarity, the unique quantum degeneracy at $\psi=\psi_\textup{U}$ and its associated spin liquid regime has not been investigated experimentally due to the scarcity of model systems.

For rare-earth triangular-lattice antiferromagnets, three classes of QSL candidate materials have been extensively studied. The first class includes the ytterbium-based compound YbMgGaO$_4$ \cite{PhysRevLett.115.167203,WOS:000391190500046,WOS:000394070700014}.  
Although a spin excitation continuum observed in
inelastic neutron scattering spectra of YbMgGaO$_4$ suggests the presence of deconfined fractionalized excitations characteristic of a QSL
\cite{WOS:000391190500046,WOS:000394070700014}, this signal could also arise from a spin glass state induced by disorder from
nonmagnetic Mg$^{2+}$ and Ga$^{3+}$ site mixing \cite{PhysRevLett.120.087201,PhysRevX.8.031028,PhysRevLett.119.157201}. The random site mixing also leads to mixed magnetic exchange interactions, resulting in broadened spin waves in a magnetic field-polarized FM phase \cite{WOS:000394070700014, WOS:000446566800007,PhysRevB.104.224433}.  Since the dispersion and energy width of spin waves in the field-induced FM state can reveal the average magnetic exchange interactions and their random distribution \cite{balents2010spin, savary2016quantum, zhou2017quantum, mila2000quantum, broholm2020quantum}, it is crucial to perform these measurements in QSL candidate materials.

The second class of triangular-lattice QSL candidate materials is the disorder-free $A$Yb$Ch_2$ ($A=$ alkali metal and $Ch=$ O, S, Se) compounds \cite{Liu_2018}. However, these materials face challenges, including susceptibility to alkali metal site deficiency and difficulties in growing high-quality single crystals \cite{WOS:000488590700022,PhysRevX.11.021044,WOS:001099033900001}. Additionally, their large AFM exchange energy scale makes it difficult to achieve a fully polarized FM state in inelastic neutron scattering experiments \cite{WOS:001070934100001,PhysRevB.109.014425}. As a consequence, 
the precise magnetic exchange interactions and the effect of alkali metal deficiency on FM spin waves of $A$Yb$Ch_2$ remain unknown.

Recently, a new class of rare-earth triangular lattice antiferromagnets, hexaaluminates \RMAO{} (RE = Pr, Nd) \cite{Ashtar2019,Bu2022,Ma2024,Cao2024,Li2024,Tu2024}, have been synthesized. Various measurements on powder \cite{Bu2022} and single crystals of \PMAO{} suggest that the system is Ising-like and may exhibit a gapless spin excitation continuum consistent with a QSL \cite{Ma2024,Cao2024,Li2024}. Single crystal X-ray diffraction has revealed the presence of quenched disorder within the mirror plane, with approximately 7\% of Pr ions displaced from their ideal positions towards the 6$h$ site \cite{Ashtar2019,Bu2022,Ma2024}. In addition, there is weak disorder between the nonmagnetic Al$^{3+}$ and Mg$^{2+}$ in \PMAO{} similar to YbMgGaO$_4$ \cite{Ashtar2019,Bu2022,Ma2024}. While the site mixing only occurs within Al/MgO$_4$ tetrahedra and is argued to have a minimal effect on mixing magnetic exchange interactions - since there is no structural disorder in AlO$_5$ and AlO$_6$ polyhedra \cite{Ashtar2019,Bu2022,Ma2024} - the magnetic exchanges and actual impact of nonmagnetic site mixing on the width of spin waves remains unknown, as there have been no inelastic neutron scattering measurements in the field-induced FM state. 

We have successfully synthesized high-quality single crystals of the Ce-based effective spin-$\frac{1}{2}$ triangular-lattice hexaaluminate \CMAO{} (Fig.~\ref{fig1}b,c). In contrast to \PMAO{}, a non-Kramers material, the ground state doublet of \CMAO{} is a Kramers doublet protected by time-reversal symmetry. While single crystal X-ray diffraction reveals the presence of quenched disorder within the mirror plane - 7\% Ce ions displaced from ideal positions - the absence of sharp transition in the zero-field heat capacity $C_{\rm p}(T)$ down to 60~mK suggests geometric frustration suppressing conventional magnetic orders (Fig.~\ref{fig2}).
By applying a 4~T $c$-axis-oriented magnetic field, we induced instrumental-energy-resolution-limited fully polarized FM spin waves, indicating that disorder in \CMAO{} has little effect on magnetic exchange interactions (Fig.~\ref{fig3}). Through the analysis of spin-wave excitations, we determine the spin Hamiltonian (Eq.~\ref{Eq:Hxxz}) featuring NN AFM interaction $J_z = 0.056(3)$~meV and FM $J_\perp=-0.024(5)$~meV. 
In zero field, \CMAO{} exhibits no static magnetic order above 100~mK.
The inelastic spin excitation spectra reveal a sharp spin-wave-like mode and a continuum of excitations (Fig.~\ref{fig4}). For energies $\hslash\omega<0.1$~meV, the excitation continuum is bounded by the sharp modes stemming from $\Gamma$ points, while for energies $\hslash\omega>0.1$~meV, the continuum scattering becomes more concentrated around the Brillouin zone boundaries. 
The neutron scattering spectrum is well reproduced by superposing an ensemble of linear spin-wave spectra from different ground states within the manifold of degenerate umbrella states (Fig.~\ref{fig5}).
Our work demonstrates that \CMAO{} represents an exactly solvable spin liquid on a triangular lattice and documents a continuum of excitations arising from the ground state degeneracy.

\section*{Experimental Results}

\begin{figure}[ht]
\centering
\includegraphics[width=0.8\textwidth]{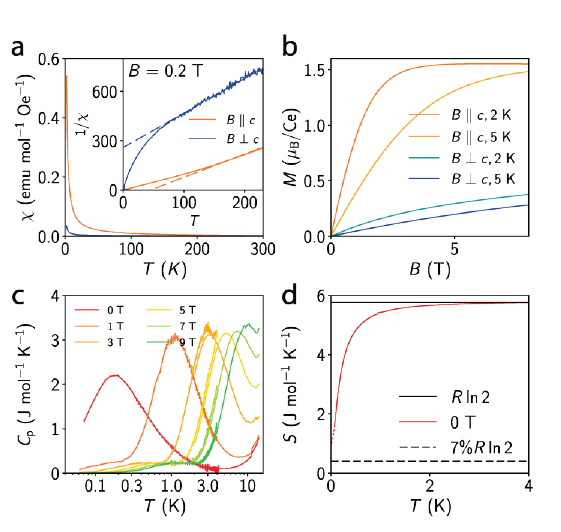}
\caption{
\textbf{Magnetic Susceptibility, Magnetization, Specific Heat Capacity, and Magnetic Entropy of \CMAO{}.}
\textbf{a}, Temperature dependence of magnetic susceptibility $\chi(T)$ for measured under an applied magnetic field of 0.2~T. The SI equivalent of emu mol$^{-1}$ Oe$^{-1}$ is 4$\pi$~cm$^3$ mol$^{-1}$. The data reveal a Curie-Weiss behavior at high temperatures and a deviation indicating magnetic correlations at lower temperatures. The inset shows the inverse susceptibility $1/\chi(T)$. \textbf{b}, Field-dependent magnetization $M(B)$ curves measured at 2~K and 5~K, parallel and perpendicular to the $c$-axis. The magnetization shows saturation around 4~T at 2~K, indicating an easy axis along the $c$-axis. \textbf{c}, Temperature and $c$-axis oriented magnetic field dependence of the heat capacity $C_\textup{p}(T)$ of \CMAO. The data show a broad peak indicative of magnetic entropy contributions. There is no contribution from the nuclear Schottky effect as all four stable isotopes of Ce carry zero nuclear spin. \textbf{d}, Entropy $S(T)$ obtained by integrating the heat capacity data. The data reach 90\% of the expected entropy release $R\ln2$ for a spin-$\frac{1}{2}$ system, after considering 7\% missing Ce ions in the triangular plane. }\label{fig2}
\end{figure}

We first investigate the magnetic properties of \CMAO{} via magnetic susceptibility, magnetization, specific heat capacity, and magnetic entropy.
Figure~\ref{fig2}a and the inset show magnetic susceptibility $\chi(T)$ and inverse susceptibility $1/\chi(T)$, respectively. Fitting $\chi(T)$ at high temperatures in magnetic fields parallel and perpendicular to the $c$-axis against Curie-Weiss behavior yields the Curie-Weiss temperature $\Theta_{\rm CW,\parallel}=45(1)$~K with an effective moment $\mu_{\rm eff,\parallel}=2.38(1)$~$\mu_{\rm B}$ and $\Theta_{CW,\perp}=-110(3)$~K with $\mu_{\rm eff,\perp}=1.90(2)$~$\mu_{\rm B}$, indicative of leading AFM coupling. $\chi$ in the $c$-axis oriented field is significantly larger than that in the in-plane field, suggesting an easy \textit{c}-axis. Figure~\ref{fig2}b presents magnetization $M(B)$ in magnetic fields parallel and perpendicular to the $c$-axis, revealing monotonic increases with the applied field. While $M(B)$ in the $c$-axis oriented field saturates below 4~T at 2~K, $M(B)$ in the in-plane field is smaller and does not saturate up to 8~T. 
Figure~\ref{fig2}c shows specific heat capacity $C_\textup{p}(T)$ in zero field, which displays a broad peak around 0.2~K, characteristic of magnetic entropy contributions and suggests the onset of coherent quantum fluctuations without a magnetic transition down to 50~mK. This broad peak shifts to higher temperatures with increasing magnetic fields along the \textit{c}-axis. Integrated magnetic entropy up to 4~K is shown in Figure~\ref{fig2}d. The entropy reaches within 90\% of $R$ln2, consistent with a spin-$\frac{1}{2}$ system \cite{WOS:000488590700021,PhysRevX.12.021015}. Note that there is no contribution from the nuclear Schottky effect as all four stable isotopes of Ce carry zero nuclear spin.

\begin{figure}[ht]
\centering
\includegraphics[width=0.8\textwidth]{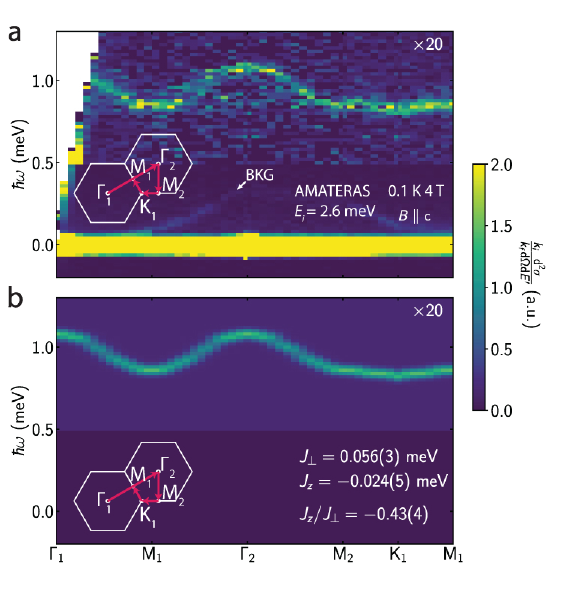}
\caption{
\textbf{Coherent Spin Waves of \CMAO{} in the Field-Polarized FM State.}
\textbf{a}, $E$-$\textbf{Q}$ spectrum under an applied magnetic field of 4~T along the $c$-axis at 0.1~K. Inset indicates the path along high-symmetry directions in the reciprocal space ($\Gamma_1 - M_1 - \Gamma_2 -M_2 - K_1 - M_1$).
The observed spectrum features sharp magnon branches, indicating that the nonmagnetic site disorder plays a negligible role in the magnetic interactions within the triangular lattice structure. The dispersive feature at low energies that does not follow the lattice symmetry is from the background of the sample environment indicated as ``BKG''. \textbf{b}, Spin-wave excitations simulated by linear spin-wave theory calculations along high-symmetry directions in a 4~T field. The parameters $J_\perp$ and $J_z$ are obtained by pixel-to-pixel fitting of the measured spectrum in panel (a) against the spin-Hamiltonian in Equation~\ref{Eq:Hxxz}. }\label{fig3}
\end{figure}

To determine the spin Hamiltonian (Eq.~\ref{Eq:Hxxz}) from the spin-wave excitations in the magnetic-field-polarized state, we carried out an inelastic neutron scattering experiment in $c$-axis-oriented magnetic fields with incident energy $E_i=2.6$~meV on AMATERAS (see Methods). Figure~\ref{fig3}a presents the spectrum versus energy and momentum along high-symmetry directions ($\Gamma_1 \rightarrow \textup{M}_1 \rightarrow \Gamma_2 \rightarrow \textup{M}_2 \rightarrow \textup{K}_1\rightarrow \textup{M}_1$) at 0.1~K in a 4~T field. 
Based on magnetization $M(B)$ data at 2~K (Fig.~\ref{fig2}b), the 4~T field is sufficient to polarize \CMAO{} into an FM state at 0.1~K.
Indeed, the spectrum exhibits a sharp spin-wave mode corresponding to magnons of well-defined energy and momentum, consistent with expectations for a field-polarized FM ground state.
Notably, in sharp contrast to the field-polarized FM state of an antiferromagnet ($0<\psi<\frac{1}{2}\pi$) in which spin waves have maximal energy at Brillouin zone boundaries, such as M and/or K points, FM spin waves of \CMAO{} have minimal energy along zone boundaries and peaks at $\Gamma$ points.
This is only possible by admitting a FM $J_z$ with an AFM $J_\perp$, placing \CMAO{} in the $\frac{1}{2}\pi<\psi<\pi$ region of the phase diagram (Fig.~\ref{fig1}a).

Furthermore, the spin-wave excitations are instrumental resolution-limited, differentiating \CMAO{} from existing spin liquid candidates \cite{balents2010spin, savary2016quantum, zhou2017quantum, mila2000quantum, broholm2020quantum}. Such sharp spin waves indicate a negligible effect of disorder and allow a precise determination of spin Hamiltonian. 
A pixel-to-pixel least squares fit of the data in Figure~\ref{fig3}a against linear spin-wave theory was conducted. The calculated spin-wave cross section was convoluted with a Gaussian function with a full-width half-maximum energy resolution of 0.07~meV (the instrumental energy resolution).
An excellent account of the data is obtained with $J_z$ = -0.024(5) meV, $J_\perp$ = 0.056(3) meV as shown in Figure~\ref{fig3}b. 
Here, $g_z=3.66$ is determined by electron spin resonance measurements (See Supplementary Materials for details).
The exchange anisotropy $J_z/J_\perp = -0.43(4)$ places \CMAO{} very close to the exactly solvable point, i.e. the boundary $\psi_\textup{U}$ between the $120^\circ$ coplanar order and the out-of-plane FM ordered phases (Fig.~\ref{fig1}a).

\begin{figure}[ht]
\centering
\includegraphics[width=1.0\textwidth]{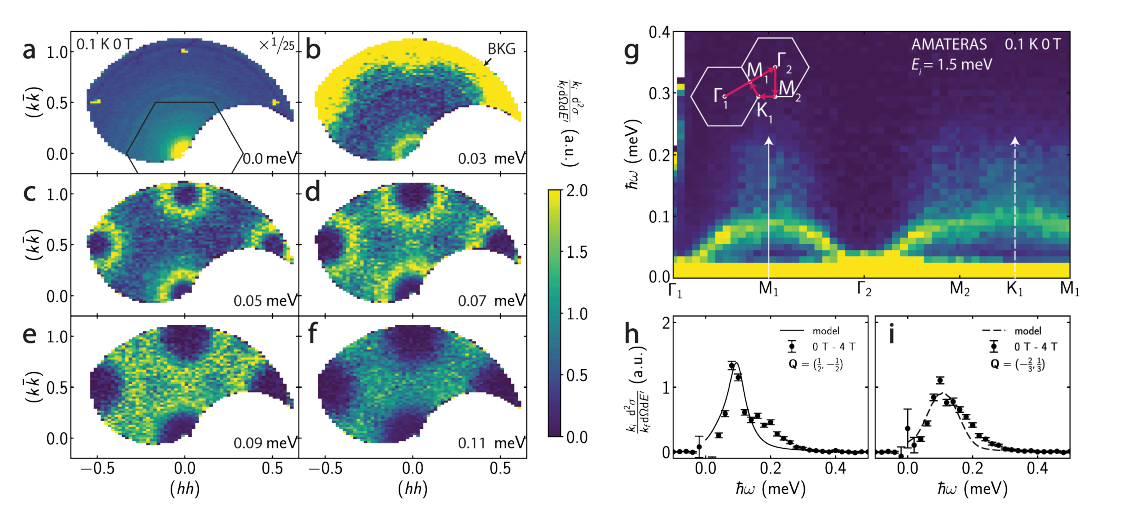}
\caption{
\textbf{Magnetic excitations of \CMAO{} in zero field. }
\textbf{a-f}, Magnetic scattering as a function of momentum and energy in the $(hk)$ plane at 0.1~K.
\textbf{g}, $E$-$\textbf{Q}$ spectrum along high-symmetry directions at 0.1~K. 
\textbf{h,i}, Constant-$\textbf{Q}$ cuts of the spectrum at M$_1$ and K$_1$ points indicated by solid and dashed arrows in panel (g). 
The solid and dashed lines are cuts at M$_1$ and K$_1$ points in the calculated spectrum presented in Figure~\ref{fig5}a.
}\label{fig4} 
\end{figure}

We now explore low-temperature magnetic excitations in zero field.
Figure~\ref{fig4}a-f presents the magnetic scattering as a function of momentum in the $(hk)$ plane with selected energies at 0.1~K. 
Consistent with the absence of anomaly in $C_\textup{p}(T)$, \CMAO{} shows no extra magnetic Bragg reflection in the elastic scattering (Fig.~\ref{fig4}a). 
As energy increases from 0.03~meV to 0.09~meV, a sharp spin-wave-like mode forming rings of scattering stems from $\Gamma$ points (Fig.~\ref{fig4}b-e), which is clear in the $E$-$\textbf{Q}$ spectrum shown in Figure~\ref{fig4}g.
Furthermore, for energies $\hslash\omega<0.1$~meV, the constant-energy spectra exhibit a continuum of excitations bounded by the sharp spin-wave-like mode around the $\Gamma$ points (Fig.~\ref{fig4}c-d). 
For energies $\hslash\omega>0.1$~meV, the continuum persists at zone boundaries up to 0.3~meV, much larger than the exchange interactions (Fig.~\ref{fig4}f-i).
Figure~\ref{fig4}h,i shows constant-$Q$ cut at M and K points, respectively. 
The continuum appears to be gapped at M and K, which distinguishes \CMAO{} from existing spin liquid candidates, such as YbMgGaO$_4$\cite{WOS:000391190500046} and NaYbSe$_2$\cite{PhysRevX.11.021044}, where continua of scattering around the Brillouin zone boundaries concentrate at $\hslash\omega=0$~meV.



\begin{figure}[ht]
\centering
\includegraphics[width=1.0\textwidth]{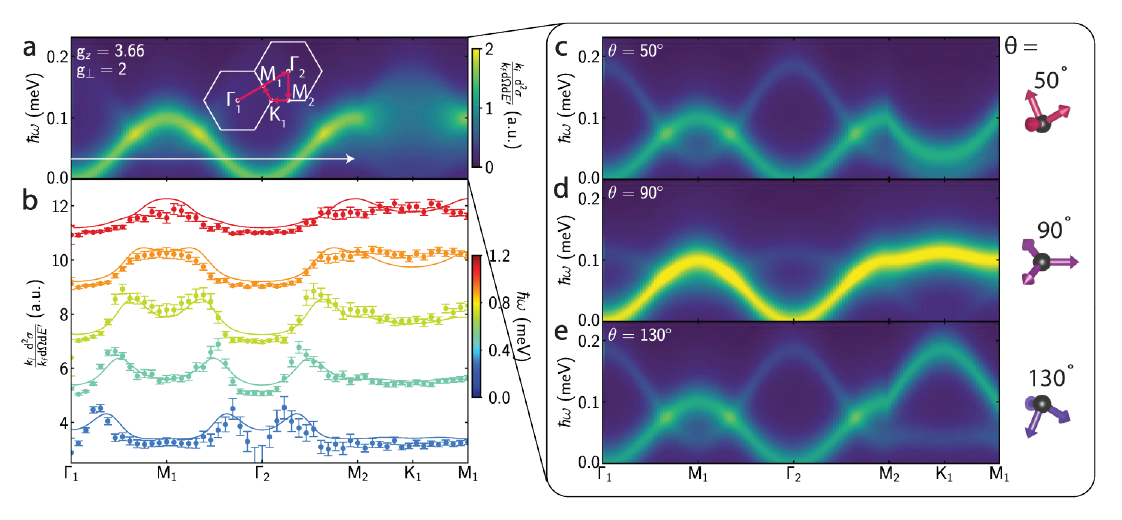}
\caption{
\textbf{Calculated Magnetic Excitations at $\psi=\psi_\textup{U}$}
\textbf{a}, Ensemble of spin-wave excitations of ground state manifolds in the umbrella state. $g_z=3.66$ is measured by ESR. Spins are lifted by $\theta$ from the triangular plane. $\theta$ is distributed as a Gaussian centered at 90$^\circ$ with 20$^\circ$ standard deviation to describe the measured data shown in Figure~\ref{fig4}.
\textbf{b}, Constant-energy cuts of measured and calculated spectra along high-symmetry directions indicated by the white arrow at various energies. 
\textbf{c-e}, Representative spin-wave excitations of ground state manifolds in the umbrella state at selected $\theta$ generated by Sunny Suite \cite{sunnysuite}.
}\label{fig5}
\end{figure}

While FM and coplanar 120$^\circ$ states exhibit sharp spin-wave excitations within the linear spin-wave theory approximation, at the exactly solvable point $\psi_\textup{U}$, continuous manifolds of ground states show exact quantum degeneracy (Fig.~\ref{fig1}a). We now consider the spectrum of the umbrella state at $\psi_\textup{U}$. Figure~\ref{fig5}a,b show spectrum and constant-energy cuts along high-symmetry directions calculated for an ensemble of spin-wave excitations of individual umbrella ground state manifolds. Representative spin-wave spectra corresponding to selected out-of-plane angle $\theta$ are shown in Figure~\ref{fig5}c-e. A quantitative agreement between measured and calculated spectra is obtained with $g_\perp=2$ and $\theta$ distributed as a Gaussian centered at $\theta=90^\circ$ with 20$^\circ$ standard deviation.
The most intense spin-wave mode persists at the same ($E$, $\textbf{Q}$) along $\Gamma$ - M, while ($E$, $\textbf{Q}$) varies dramatically at the Brillouin boundaries M - K, resulting in a continuum of spin excitations.

\section*{Discussion}

To understand the novel state of matter such as a QSL, it is essential to 
identify the microscopic Hamiltonian responsible for the observed exotic behavior. For the $S=\frac{1}{2}$ 2D honeycomb lattice, Kitaev’s exactly solvable model with bond-dependent NN Ising interactions has a QSL ground state, where the excitations are itinerant Majorana fermions and static $Z_2$ fluxes useful for fault-tolerant quantum computation \cite{KITAEV20032,KITAEV20062}. Although many candidate materials exist, there is no confirmed case of a Kitaev QSL material \cite{WOS:000540317900009}. For 2D kagome lattice QSL candidate materials, the effect of magnetic and nonmagnetic disorders may form a spin glass state that mimics the signature of a QSL \cite{WOS:000312488200051,RevModPhys.88.041002}. While recent progress on triangular lattice materials is encouraging \cite{WOS:001070934100001,PhysRevB.109.014425,WOS:000989944200002}, a fundamental limitation is the inability to determine the magnetic exchange interactions through FM polarized spin wave measurements.  \CMAO{} is the rare case where field-induced FM spin waves are resolution-limited, ruling out that nonmagnetic site disorder has a large impact on the spin Hamiltonian (Fig.~\ref{fig3}).  By determining the magnetic exchange interactions in \CMAO{}, we find that the system is near the boundary between 
the $120^\circ$ coplanar order and the out-of-plane FM-ordered phases 
of the XXZ model (Fig.~\ref{fig1}a). These results are consistent with spin excitation spectra observed at zero magnetic fields, indicating that the observed continuum-like excitations do not arise from fractionalized spinon excitations of a QSL but from degenerate umbrella ground state manifold. Therefore, our work provides experimental realization of an exactly solvable spin liquid on a triangular lattice and documents a continuum of excitations due to the ground state degeneracy of the $120^\circ$ umbrella order.

\section*{Methods}

\subsection*{Crystal Growth}

Polycrystalline \CMAO{} samples were synthesized through a solid-state reaction, where CeO$_2$, MgO, and Al$_2$O$_3$ were mixed in precise stoichiometric ratios, ground, and pelletized for uniformity. The pellets were then calcined at 1250$^\circ$C to 1550$^\circ$C for 72 hours, with intermediate grinding to improve purity and crystallinity. Single crystals were grown using the laser diode floating zone method, yielding high-quality \CMAO{} crystals with well-defined $ab$-plane facets, suitable for magnetic and neutron scattering studies.

\subsection*{X-ray Diffraction}

Single-crystal X-ray diffraction data for \CMAO{} were collected at room temperature using a Bruker Kappa D8 Quest diffractometer with Mo K$\alpha$ radiation. The structure was solved and refined using standard crystallographic software, revealing that \CMAO{} crystallizes in the hexagonal $P6_3/mmc$ space group with lattice parameters $a = 5.5813(5) \text{\AA}$ and $c = 21.904(2) \text{\AA}$. The structure consists of CeO$_{12}$ polyhedra, Mg/AlO$_{4}$ tetrahedra, and AlO$_{6}$ octahedra, typical of magnetoplumbite-type structures. Ce$^{3+}$ ions occupy specific Wyckoff positions, forming a distorted anti-cuboctahedral environment, while Mg$^{2+}$ and Al$^{3+}$ ions share positions, leading to partial occupancy and substitutional disorder that cannot be fully resolved by X-ray diffraction alone, necessitating neutron diffraction for differentiation. See Supplementary Materials for details.

\subsection*{Neutron Diffraction}

Neutron diffraction data for \CMAO{} were collected at room temperature using the TOPAZ instrument at the Spallation Neutron Source (SNS) with neutron wavelengths of 0.4 - 3.5 ~\text{\AA}. The measurements employed time-of-flight (TOF) methods, yielding wavelength-resolved Laue patterns. Absorption corrections were applied using the multi-scan method. The crystal structure was determined to be in the hexagonal $P6_3/mmc$ space group, with lattice parameters $a = 5.5949(3) \text{\AA}$ and $c = 21.9286(19) \text{\AA}$, and a unit cell volume of $V = 594.46(7) \text{\AA}^3$. The structure was refined using the JANA2020 software package, and the Ce deficiency in \CMAO{} was found to be approximately 7\%, similar to the Pr deficiency in \PMAO{} \cite{Cao2024}. See Supplementary Materials for details.

\subsection*{Magnetic Susceptibility and Heat Capacity}

Magnetic susceptibilities were measured using a Quantum Design Magnetic Property Measurement System (MPMS). An oriented piece of \CMAO{} crystal was mounted on an MPMS sample holder using GE varnish. Heat capacity data were collected in a Quantum Design Physical Property Measurement System (PPMS). An oriented piece of \CMAO{} crystal was mounted on the heat capacity puck using Apiezon N Grease. The addenda heat capacity was previously measured and subtracted. Heat capacity data below 1.8 K were collected with a dilution refrigerator.

\subsection*{Inelastic Neutron Scattering}

In the inelastic neutron scattering experiment on \CMAO{}  conducted at the cold-neutron disk-chopper spectrometer (AMATERAS) \cite{Nakajima2011} Japan Proton Accelerator Research Complex (J-PARC), a single-piece of single crystalline \CMAO{} was aligned in the $(hk0)$ scattering plane.  We define the momentum transfer $\textbf{Q}$ in three-dimensional reciprocal space in \AA$^{-1}$ as $\textbf{Q} = h\textbf{a}^* + k\textbf{b}^* + l\textbf{c}^*$, where $h$, $k$, and $l$ are Miller indices and $\textbf{a}^* = \hat{a}2\pi/a$, $\textbf{b}^* = \hat{b}2\pi/b$, and $\textbf{c}^* = \hat{c}2\pi/c$ with $a = b = 5.5813(5)  \text{\AA}$, $c = 21.904(2)$~$\text{\AA}$  in the $P6_3/mmc$ space group. For time-of-flight neutron scattering experiments on AMATERAS, incident neutron energies of $E_i = 1.5$ and 2.6~meV were used with instrumental energy resolution at elastic positions of 0.037 and 0.075 meV, respectively. We aligned the sample in the $(hh0) \times (k\bar{k}0)$ scattering plane with 0 and 4~T vertical field along the $(001)$ direction at 100~mK.

\bibliography{main_ref}

\noindent\textbf{Acknowledgements}\\
We gratefully acknowledge valuable discussions with H. Hu, J. Zhang, S. Zhang, Y. Gao, and Shiyan Li. 
The neutron scattering work at Rice was supported US DOE BES DE-SC0012311 (P.D.). 
The single crystal growth work was supported by the Robert A. Welch Foundation under Grant No. C-1839 (P.D.). 
Crystal growth by B.G., X.X., and S.W.C. at Rutgers was supported by the visitor program at the center for Quantum Materials Synthesis (cQMS), funded by the Gordon and Betty Moore Foundation’s EPiQS initiative through grant GBMF6402, and by Rutgers University. 
The theoretical work done by C.L. and L.B. was supported by the DOE, Office of Science, Basic Energy Sciences under Award No. DE-FG02-08ER46524 and the Simons Collaboration on Ultra-Quantum Matter. 
C.L. acknowledges the fellowship support from the Gordon and Betty Moore Foundation through the Emergent Phenomena in Quantum Systems (EPiQS) program.
Z.M. acknowledges the National Natural Science Foundation of China with Grant No. 12204160.
C.W. acknowledges the National Research Foundation of Korea (NRF), Ministry of Science and ICT (No. 2022M3H4A1A04074153).
G.T.M. and J.Y.C. acknowledge Welch Foundation, United States AA-2056-20240404.
The neutron scattering experiment at the MLF of J-PARC was performed under proposal No. 2022B0242.
This research used resources at the Spallation Neutron Source, a DOE Office of Science User Facility operated by Oak Ridge National Laboratory.
\\

\noindent\textbf{Author contributions}\\
T.C., Z.M., and P.D. initiated this work. B.G., X.X., and S.W.C. prepared the samples. X.X. and C.W. measured magnetic susceptibility and heat capacity. B.G., T.C., P.D., M.L.K., S.O.K., and N.M. conducted neutron scattering experiments. S.J.M., J.T.R., and X.H. measured ESR. G.T.M. and J.Y.C. performed X-ray diffraction. X.W. performed the neutron diffraction. B.G., T.C., C.L., S.Z., O.T., L.B., and P.D. wrote the manuscript with input from all coauthors.\\

\noindent\textbf{Competing interests}\\
The authors declare no competing interests. Certain equipment, instruments, software, or materials are identified in this paper in order to specify the experimental procedure adequately.  Such identification is not intended to imply recommendation or endorsement of any product or service by NIST, nor is it intended to imply that the materials or equipment identified are necessarily the best available for the purpose.

\section*{Figures}

\end{document}


\preprint{APS/123-QED}

\title{Supplementary Materials for Continuum of Spin Excitations in an Exactly Solvable Spin Liquid \CMAO}

\author{Bin Gao}
\thanks{These authors contributed equally to this work.}

\author{Tong Chen}
\thanks{These authors contributed equally to this work.}

\author{Chunxiao Liu}
\thanks{These authors contributed equally to this work.}

\author{Mason L. Klemm}
\author{Shu Zhang}
\author{Zhen Ma}
\email{zma@hbnu.edu.cn}
\author{Xianghan Xu}
\author{Choongjae Won}
\author{Gregory T. McCandless}
\author{Naoki Murai}
\author{Seiko Ohira-Kawamura}
\author{Stephen J. Moxim}
\author{Jason T. Ryan}
\author{Xiaozhou Huang}
\author{Xiaoping Wang}
\author{Julia Y. Chan}
\author{Sang-Wook Cheong}
\author{Oleg Tchernyshyov}
\author{Leon Balents}
\email{balents@spinsandelectrons.com}
\author{Pengcheng Dai}
\email{pdai@rice.edu}

\date{\today}
\maketitle


\section{\label{appendix: Crystal Growth} Crystal Growth}

Polycrystalline \CMAO{} samples were synthesized using a standard solid-state reaction technique. High-purity starting materials - CeO$_2$, MgO, and Al$_2$O$_3$ - were accurately weighed and mixed to achieve the desired stoichiometric ratios. The mixture was repeatedly ground and pelletized to ensure homogeneity, then placed in alumina crucibles and calcined initially at 1250$^\circ$C and finally at 1550$^\circ$C for approximately 72 hours. Intermediate grinding and re-pelletizing were conducted between calcination steps to enhance purity and crystallinity. Powder X-ray diffraction confirmed the phase purity of the synthesized samples, verifying the formation of single-phase \CMAO{} without detectable impurities.

Single crystals were grown using the laser diode floating zone (LFZ) method in a forming gas flow (8\% {H$_2$} in Ar), producing \CMAO{} crystals with well-defined $ab$-plane facets and high structural quality, suitable for detailed magnetic and neutron scattering experiments.

\begin{figure}[ht]
\centering
\includegraphics[width=0.9\textwidth]{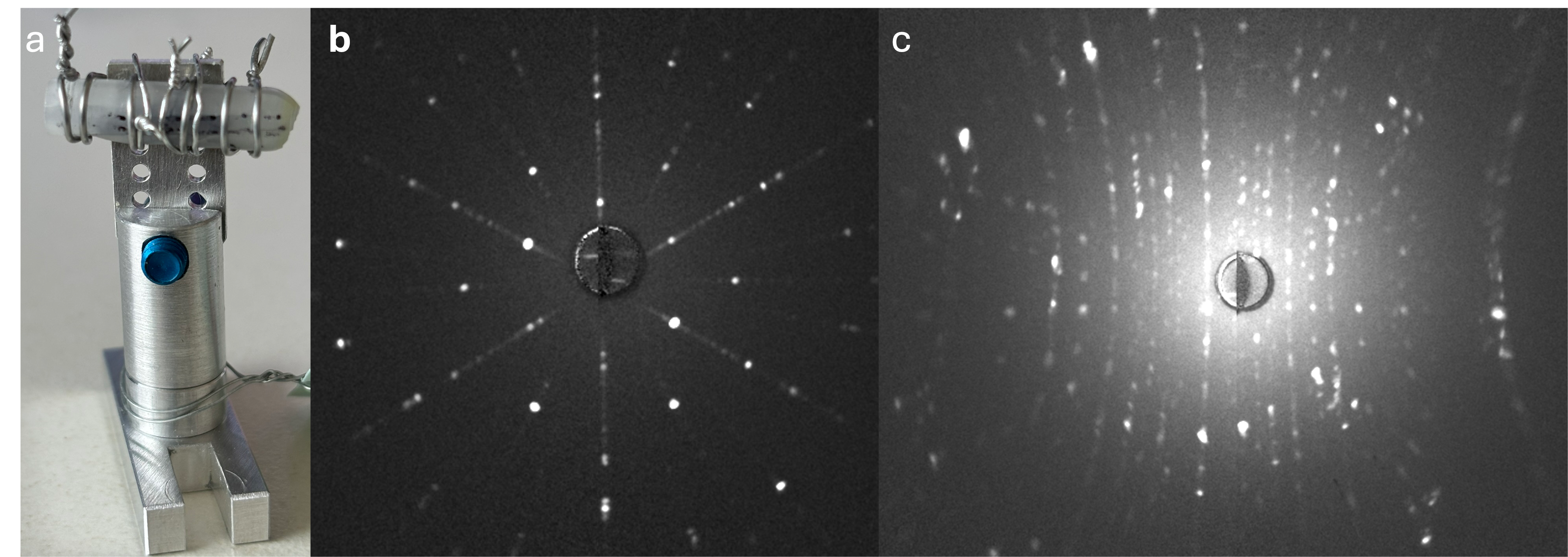}
\caption{a, Photo of the single crystalline \CMAO{} about 2 grams aligned in the $[h,k,0]$ scattering plane. b, X-ray Laue pattern in the $[0,0,1]$ direction. c, X-ray Laue pattern in the $[1,0,0]$ direction.}\label{figS1}
\end{figure}


\section{\label{appendix: Single Crystal Neutron Diffraction} Single Crystal Neutron Diffraction}

Neutron diffraction data were collected at room temperature using the TOPAZ instrument at the Spallation Neutron Source (SNS). The measurements were performed with neutron wavelengths of 0.4 \text{\AA} - 3.5 \text{\AA}. The sample was mounted on a goniometer and data were collected using time-of-flight (TOF) methods, which provided wavelength-resolved Laue patterns. The data were corrected for absorption effects using the multi-scan method, with Tmin = 0.943 and Tmax = 0.958.

The crystal structure of \CMAO{} was determined using neutron diffraction data. The compound crystallizes in the hexagonal space group $P6_3/mmc$ with lattice parameters $a = 5.5949(3) \text{\AA}$ and $c = 21.9286(19) \text{\AA}$, resulting in a unit cell volume of $V = 594.46(7) \text{\AA}^3$. The refinement was carried out using the JANA2020 software package. The final refinement statistics are summarized in Table \ref{Tab1} and the fractional atomic coordinates and isotropic displacement parameters of the neutron diffraction experiment are listed in Table \ref{Tab2}. The percentage of Ce deficiency ($7\%$) is similar to the Pr deficiency in \PMAO{} \cite{Cao2024}.

\begin{table}[ht]
\centering
\caption{Neutron Experiment Details}
\begin{tabular}{l l}
\hline
Crystal data & \\ \hline
Chemical formula & Ce\(_{0.931}\)Mg\(_{0.801}\)Al\(_{11.199}\)O\(_{19}\) \\
Mr & 756.1 \\
Crystal system space group & Hexagonal P6\(_3\)/mmc \\
Temperature (K) & 293 \\
a, c (Å) & 5.5949 (3), 21.9286 (19) \\
V (Å\(^3\)) & 594.46 (7) \\
Z & 2 \\
Radiation type & Neutrons, \(\lambda = 0.4 - 3.5\) Å \\
µ (mm\(^{-1}\)) & 0.0351 + 0.0006\(\lambda\) \\
Crystal size (mm) & 1.85 × 1.55 × 1.10 \\
Data collection & \\ \hline
Diffractometer & TOPAZ at SNS \\
T\(_{min}\), T\(_{max}\) & 0.943, 0.958 \\
No. of measured, independent, and observed [I $>$ 2\(\sigma\)(I)] reflections & 9109, 9064, 9057 \\
R\(_{int}\) & 0.158 \\
(sin \(\theta\)/\(\lambda\))\(_{max}\) (Å\(^{-1}\)) & 1.269 \\
Refinement & \\ \hline
R[F $>$ 3\(\sigma\)(F)], wR(F), S & 0.035, 0.097, 1.15 \\
No. of reflections & 9064 \\
No. of parameters & 64 \\
\(\Delta\rho_{max}\), \(\Delta\rho_{min}\) (f Å\(^{-3}\)) & 0.31, -0.68 \\
Computer programs & SNS EPICS, Mantid, python program, JANA2020 \\
\hline
\end{tabular}\label{Tab2}
\end{table}

\begin{table}[ht]
\centering
\caption{Fractional atomic coordinates and isotropic displacement parameter (Å\(^2\)) from neutron diffraction}
\begin{tabular}{c c c c c c c c c}
\hline
\textbf{Atom} & \textbf{Label} & \textbf{x} & \textbf{y} & \textbf{z} & \textbf{Occ.} & \textbf{U} & \textbf{Site} & \textbf{Sym.} \\
\hline
Ce  & Ce1 & 2/3 & 1/3 & 1/4 & 0.931 & 0.005 & 2d & -6m2 \\
Al  & Al1 & 0 & 0 & 0 & 1 & 0 & 2a & -3m. \\
Al  & Al2 & 0 & 0 & 0.24170 & 0.5 & 0.005 & 4e & 3m. \\
Mg  & Mg3 & 1/3 & 2/3 & 0.02752 & 0.401 & 0.001 & 4f & 3m. \\
Al  & Al3 & 1/3 & 2/3 & 0.02752 & 0.599 & 0.001 & 4f & 3m. \\
Al  & Al4 & -0.16753 & -0.33555 & 0.10823 & 1 & 0.001 & 12k & m. \\
Al  & Al5 & 1/3 & 2/3 & 0.18991 & 1 & 0.002 & 4f & 3m. \\
O   & O1  & 0 & 0 & 0.15122 & 1 & 0.002 & 4e & 3m. \\
O   & O2  & 2/3 & 1/3 & 0.05788 & 1 & 0.002 & 4f & 3m. \\
O   & O3  & 0.18119 & 0.36238 & 1/4 & 1 & 0.004 & 6h & mm2 \\
O   & O4  & 0.15241 & 0.30482 & 0.05355 & 1 & 0.004 & 12k & m. \\
O   & O5  & 0.50507 & 1.01014 & 0.15131 & 1 & 0.002 & 12k & m. \\
\hline
\end{tabular}\label{Tab1}
\end{table}


\section{\label{appendix: Single Crystal X-ray Diffraction} Single Crystal X-ray Diffraction}

\begin{figure}[ht]
\centering
\includegraphics[width=0.65\textwidth]{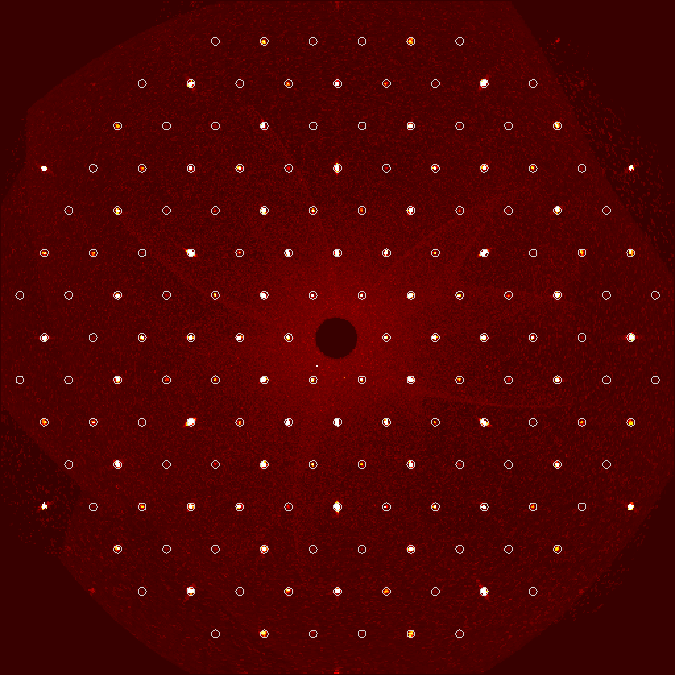}
\caption{The $[h,k,0]$ precession images (with an overlay of indexing circles) from the single crystalline X-ray diffraction of \CMAO{} }\label{figXRDHK0}
\end{figure}

\begin{figure}[ht]
\centering
\includegraphics[width=0.65\textwidth]{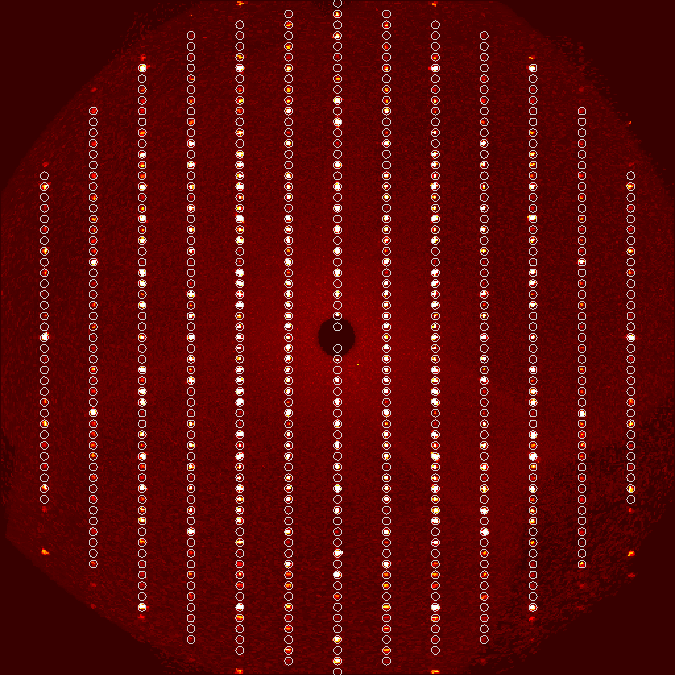}
\caption{The $[h,0,l]$ precession images (with an overlay of indexing circles) from the single crystalline X-ray diffraction of \CMAO{} }\label{figXRDH0L}
\end{figure}

Single-crystal X-ray diffraction data were collected at room temperature using a Bruker Kappa D8 Quest diffractometer equipped with a Photon 100 CMOS detector and a Mo K$\alpha$ ($\lambda = 0.71073 \textup{\AA}^3)$ radiation source. Data collection was carried out using $\omega$ and $\phi$ scan modes with frame widths of 0.5\textdegree. The collected data were processed using the Bruker APEX4 software suite. Absorption corrections were applied using the SADABS program. Fig \ref{figXRDHK0} and Fig \ref{figXRDH0L} show the precession images in the $[h,k,0]$ and $[h,0,l]$ planes.

\begin{table}[ht]
\centering
\caption{Experimental details of single crystal X-ray diffraction}
\begin{tabular}{ll}
\hline
\textbf{Crystal data} & \\
Chemical formula & Ce\textsubscript{0.94}Mg\textsubscript{0.8}Al\textsubscript{11.2}O\textsubscript{19} \\
$M_r$ & 758.04 \\
Crystal system, space group & Hexagonal, \textit{P}6\textsubscript{3}/\textit{mmc} \\
Temperature (K) & 298 \\
$a, c$ (\AA) & 5.5813(5), 21.904(2) \\
$V$ (\AA$^3$) & 590.91(12) \\
$Z$ & 2 \\
Radiation type & Mo K$\alpha$ \\
$\mu$ (mm$^{-1}$) & 4.67 \\
Crystal size (mm) & 0.04 $\times$ 0.04 $\times$ 0.04 \\
\hline
\textbf{Data collection} & \\
Diffractometer & Bruker Kappa D8 Quest CPAD \\
Absorption correction & Multi-scan SADABS \\
$T_{\min}, T_{\max}$ & 0.654, 0.736 \\
No. of measured, independent, and observed [I $>$ 2$\sigma$(I)] reflections & 25075, 392, 360 \\
$R_{\text{int}}$ & 0.071 \\
(sin $\theta/\lambda$)$_{\text{max}}$ (\AA$^{-1}$) & 0.713 \\
\hline
\textbf{Refinement} & \\
$R[F^2 > 2\sigma(F^2)], wR(F^2), S$ & 0.027, 0.073, 1.26 \\
No. of reflections & 392 \\
No. of parameters & 43 \\
$\Delta \rho_{\text{max}}, \Delta \rho_{\text{min}}$ (e \AA$^{-3}$) & 1.11, -1.42 \\
\hline
\end{tabular}
\end{table}

\begin{table}[ht]
\centering
\caption{Fractional atomic coordinates and isotropic displacement parameters (\AA$^2$)}
\begin{tabular}{cccccc}
\hline
Atom & $x$ & $y$ & $z$ & $U_{\text{iso}}/U_{\text{eq}}$ & Occupancy \\
\hline
Ce1 & 2/3 & 1/3 & 1/4 & 0.01006(19) & 0.947(5) \\
Al1 & 0 & 0 & 0 & 0.0050(5) & 1 \\
Al2 & 0 & 0 & 0.2420(6) & 0.009(2) & 0.5 \\
Mg3 & 1/3 & 2/3 & 0.02730(9) & 0.0050(4) & 0.401 \\
Al3 & 1/3 & 2/3 & 0.02730(9) & 0.0050(4) & 0.599 \\
Al4 & -0.16753(10) & -0.33506(19) & 0.10821(5) & 0.0046(3) & 1 \\
Al5 & 1/3 & 2/3 & 0.18988(8) & 0.0046(4) & 1 \\
O1 & 0 & 0 & 0.15110(19) & 0.0074(8) & 1 \\
O2 & 2/3 & 1/3 & 0.0578(2) & 0.0066(8) & 1 \\
O3 & 0.1801(4) & 0.3602(7) & 1/4 & 0.0088(7) & 1 \\
O4 & 0.1527(3) & 0.3055(5) & 0.05355(11) & 0.0081(5)
 & 1 \\
O5 & 0.5044(2) & 1.0088(5) & 0.15124(10) & 0.0050(5) & 1 \\
\hline
\end{tabular}
\end{table}

\begin{table}[ht]
\centering
\caption{Atomic displacement parameters for \CMAO{}}
\begin{tabular}{cccccccc}
\hline
Atom & $U^{11}$ & $U^{22}$ & $U^{33}$ & $U^{12}$ & $U^{13}$ & $U^{23}$ \\
\hline
Ce1  & 0.0116 (2) & 0.0116 (2) & 0.0073 (2) & 0.00579 (11) & 0 & 0 \\
Al1  & 0.0051 (7) & 0.0051 (7) & 0.0048 (10) & 0.0025 (3) & 0 & 0 \\
Al2  & 0.0039 (8) & 0.0039 (8) & 0.020 (7) & 0.0020 (4) & 0 & 0 \\
Mg3  & 0.0044 (6) & 0.0044 (6) & 0.0061 (8) & 0.0022 (3) & 0 & 0 \\
Al3  & 0.0044 (6) & 0.0044 (6) & 0.0061 (8) & 0.0022 (3) & 0 & 0 \\
Al4  & 0.0040 (4) & 0.0041 (5) & 0.0057 (5) & 0.0021 (3) & 0.00001 (15) & 0.0000 (3) \\
Al5  & 0.0047 (5) & 0.0047 (5) & 0.0052 (6) & 0.0023 (3) & 0 & 0 \\
O1   & 0.0070 (12) & 0.0070 (12) & 0.0082 (17) & 0.0035 (6) & 0 & 0 \\
O2   & 0.0060 (12) & 0.0060 (12) & 0.0081 (19) & 0.0030 (6) & 0 & 0 \\
O3   & 0.0115 (13) & 0.0082 (16) & 0.0055 (14) & 0.0041 (8) & 0 & 0 \\
O4   & 0.0065 (9) & 0.0109 (11) & 0.0084 (11) & 0.0055 (6) & -0.0012 (4) & -0.0025 (8) \\
O5   & 0.0049 (8) & 0.0062 (11) & 0.0042 (10) & 0.0031 (5) & 0.0005 (4) & 0.0009 (7) \\
\hline
\end{tabular}
\end{table}

The crystal structure was solved by intrinsic phasing methods and refined using full-matrix least-squares techniques on $F^2$ with SHELXT and SHELXL programs. \CMAO{} crystallizes in the hexagonal $P6_3/mmc$ space group with lattice parameters $a = 5.5813(5)  \text{\AA}$, $c = 21.904(2) \text{\AA}$, and $V = 590.91(12) \text{\AA}^3$. The structure is composed of CeO$_{12}$ polyhedra, Mg/AlO$_{4}$ tetrahedra, and AlO$_{6}$ octahedra, forming a layered arrangement typical of magnetoplumbite-type structures. The Ce$^{3+}$ ions occupy the 2d Wyckoff positions and are coordinated by twelve oxygen atoms forming a distorted anti-cuboctahedral environment. The Mg$^{2+}$ and Al$^{3+}$ ions share the 4f Wyckoff position, leading to partial occupancy and substitutional disorder. Mg and Al cannot be distinguished by X-ray diffraction, which can only be solved by neutron diffraction. The model used in X-ray refinement was further refined by the model from the neutron diffraction experiment.

\section{\label{appendix: Electron Spin Resonance } Electron Spin Resonance}

We performed electron spin resonance (ESR) measurement on a fragment of the \CMAO{} crystal used in the neutron experiments using a commercially available, X-band ESR spectrometer. The magnetic field was oriented parallel to the material $c$-axis ($\pm$ 2$^\circ$). The measurement temperature was at 5~K, and the microwave frequency was 9.48858~GHz. The dominating feature in the ESR response is a single line with a $g_z$ of 3.66(5). Additionally, small ESR features are observed at magnetic fields between 250~mT and 450~mT.

\begin{figure}[ht]
\centering
\includegraphics[width=1.0\textwidth]{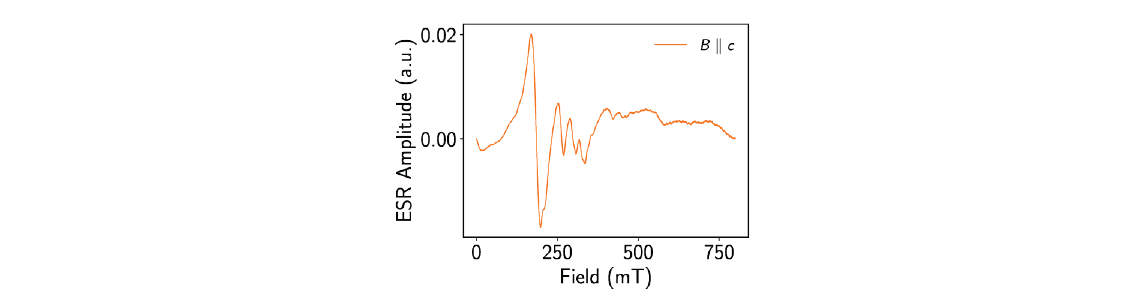}
\caption{ESR spectrum taken at 5~K with the applied magnetic field oriented parallel to the material $c$-axis.}\label{figS2}
\end{figure}


\section{\label{appendix: Theoretical Model} Theoretical Model}

\subsection{\label{appendix: Symmetry-allowed Hamiltonian} Symmetry-allowed Hamiltonian}

\CMAO{} crystallizes in space group $P6_3/mmc$ (No. 194). The Ce$^{3+}$ ions sit at mirror plane $m_z$. Point group is $D_{6h}=\langle m_z, m_x, i, c_3\rangle=\langle m_x, i, s_2, c_3\rangle$, where $s_2$ is twofold screw. Note that the mirror $m_z$ contains the Ce$^{3+}$ ions, and the center of the inversion $i$ is at the midpoint between two Ce$^{3+}$ ions. This should be Wyckoff position 2$c$, with site symmetry $\bar{6}m2=D_{3h}$ and coordinates (1/3, 2/3, 1/4) and (2/3, 1/3, 3/4). Each site is symmetric under $D_{3h}=\langle m_z, m_x, c_3 \rangle = \langle c_3, c'_2, m_z \rangle$, where $c'_2 = m_z m_x$ is a twofold axis along $y$, but not $s_2$ or $i$.

We first write down the symmetry allowed Hamiltonian. For the NN bond along (100), the tensor $J^{\mu\nu}$ in the exchange term $S_i^\mu J^{\mu\nu} S_j^\nu, \mu,\nu=1,2,3$ is subject to the following symmetry constraints: the NN bond is mapped back to itself under $m_z$: $S \rightarrow (-S^x, -S^y, S^z)$, giving $J_{13}=J_{23}=0$; under $m_x$: $S \rightarrow (S^x, -S^y, -S^z)$, giving $J_{12}=J_{13}=0$. These constraints the most general exchange model at NN level to be:
\begin{equation}
{\cal H}_{NN} = \sum_{<i,j>} [J_z  S_i^z S_j^z + J_\perp (S_i^x S_j^x + S_i^y S_j^y) + J_c(\textbf{\textit{f}}_{ij}\cdot\textbf{\textit{S}}_i)(\textbf{\textit{f}}_{ij}\cdot\textbf{\textit{S}}_j) + D_{ij}\hat{\textbf{\textit{z}}}\cdot\textbf{\textit{S}}_i \times \textbf{\textit{S}}_j],
\label{Eq:H1}
\end{equation}
where the first two terms constitute the usual XXZ model. The third term breaks the U(1) symmetry of the XXZ model to discrete symmetry, where the $\textbf{\textit{f}}_{ij}$ is the unit vector along the NN bond $\langle ij \rangle$. The DM vector $\textbf{\textit{D}}_{ij}=D_{ij}\hat{\textbf{\textit{z}}}=D\hat{\textbf{\textit{z}}}$ for $i\rightarrow j$ along +0$^\circ$, +120$^\circ$, and +240$^\circ$. Such a DM vector is the only one allowed by two perpendicular mirrors $m_z$ and $m_x$. In the following discussion, we will focus on this NN exchange model with $D=0$, since the DM interaction tends to introduce incommensurate effects \cite{shan2021torque, el2022frustrated} and split the spin waves, which are not observed in the experiments. 

We reparameterize the symmetry-allowed Hamiltonian as 
\begin{equation}
{\cal H}_{NN} = \sum_{<i,j>} [J_z  S_i^z S_j^z + J_\pm (S_i^+ S_j^- + S_i^- S_j^+)+J_{\pm\pm}(\gamma_{ij}S_i^+S_j^+ + \gamma_{ij}^* S_i^- S_j^-)],
\label{Eq:H2}
\end{equation}
where $\gamma_{ij}=1, e^{i \frac{2\pi}{3}}, e^{-i \frac{2\pi}{3}}$ for $\langle ij \rangle$ along the NN bonds, and $S^\pm=S^x \pm iS^y$. The parameters $J_\pm$ and $J_{\pm\pm}$ are related to those in Equation~\ref{Eq:H1} by $J_\pm=J_\perp/2+J_c/4$ and $J_{\pm\pm}=J_c/4$.

\subsection{\label{appendix: Exchange Parameters} Exchange Parameters}

In the main text, we fit the inelastic neutron scattering data in a 4~T $c$-axis-oriented magnetic field against the simplified XXZ model. Here, we further investigate the full symmetry-allowed Hamiltonian described by Equation~\ref{Eq:H2}. It is known that the magnon dispersion has the form \cite{paddison2017continuous}:

\begin{equation}
E(\textbf{\textit{q}})=\sqrt{(g_z \mu_\textup{B} B - 3 J_z + J_\pm f(\textbf{\textit{q}}))^2 - |J_{\pm\pm} g(\textbf{\textit{q}})|^2},
\end{equation}
where $f$ and $g$ are two form factors
\begin{equation}
f(\textbf{\textit{q}})=\sum_{i=1}^6 \cos (\textbf{\textit{q}}\cdot\textbf{\textit{r}}_i), \quad g(\textbf{\textit{q}})=\sum_{i=1}^6 \gamma_i^* \cos (\textbf{\textit{q}}\cdot\textbf{\textit{r}}_i).
\end{equation}

We use the dispersion at three high-symmetry momenta, $\Gamma$, K, and M, to extract the three exchange parameters $(J_z,J_\pm,J_{\pm\pm})$. Denote the magnon excitation energy at these momenta as $E_\Gamma$, $E_\textup{K}$, and $E_\textup{M}$, we have
\begin{equation}
J_z = \frac{1}{9}(3Bg_z \mu_\textup{B} - E_\Gamma - 2E_\textup{K}), \quad
J_\pm = \frac{1}{9}(E_\Gamma-E_\textup{K}), \quad
J_{\pm\pm}=\frac{1}{36} \sqrt{E_\Gamma^2 + 64 E_\textup{K}^2 + 16 E_\Gamma E_\textup{K} - 81E_\textup{M}^2}.
\end{equation}

Taking the following value from the neutron data
\begin{equation}
E_\Gamma=1.084~\textup{meV}, \quad E_\textup{K} = 0.847~\textup{meV}, \quad E_\textup{M}=0.873~\textup{meV},
\end{equation}
and use $g_z=3.66$ from ESR experiment, we obtain
\begin{equation}
J_z = -0.0262~\textup{meV}, \quad J_\pm = 0.0263~\textup{meV}, \quad J_{\pm\pm} = 0.0060~\textup{meV}.
\label{Eq:H2p}
\end{equation}

We plot the classical phase diagram for the $(J_z,J_\pm,J_{\pm\pm})$ model. We note that, by admitting $J_{\pm\pm}$, the parameters place \CMAO{} right at the boundary of the coplanar 120$^\circ$ and the FM ordered phases. It is a tri-critical point among the two phases and strip $y$ phase.

\begin{figure}[ht]
\centering
\includegraphics[width=0.45\textwidth]{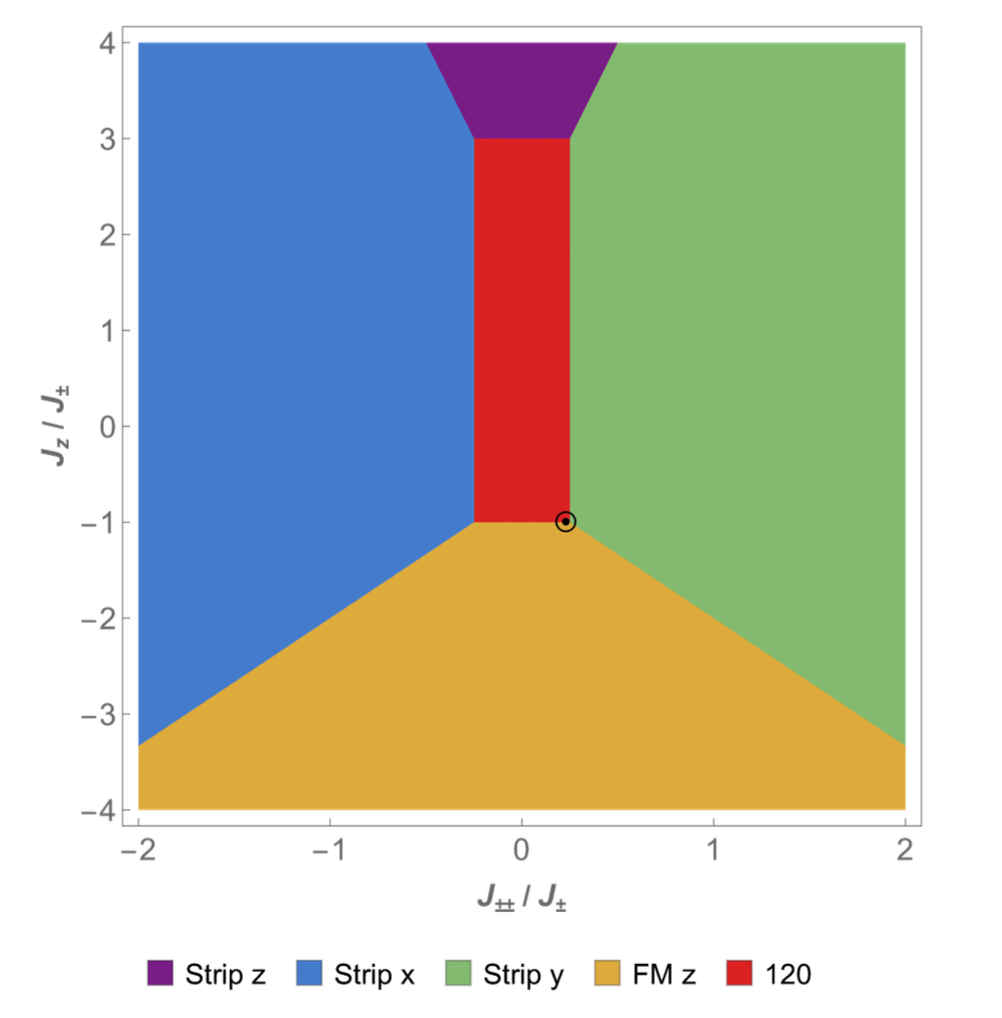}
\caption{Classical phase diagram for the $(J_z,J_\pm,J_{\pm\pm})$ model (Eq.~\ref{Eq:H2}). We assumed $J_\pm>0$. The strip $x,y,z$ phases are Neel ordered phases with spin along $x,y,z$; the FM $z$ denote the ferromagnetic phase with spins polarized along $z$; the 120$^\circ$ phase denotes the coplanar 120$^\circ$ order. The black dot indicates the parameters for \CMAO{}.
}
\label{figS4}
\end{figure}

\subsection{\label{appendix: The psiU model } The $\psi_\textup{U}$ Model ($J_z=-J_\pm$)}

Classical magnetic order at $J_z=-J_\pm$ has been studied by Miyashita \cite{miyashita1986ground}. The ground state is a three-sublattice umbrella order of the following form
\begin{equation}
\textbf{\textit{S}}_d = (\sin \theta \cos (\phi+2\pi d/3), \sin \theta \sin (\phi + 2\pi d /3), \cos \theta), \quad d=0,1,2\equiv a,b,c,
\end{equation}
where we use $d=0,1,2 \equiv a,b,c$ to label the three sublattices. Note that when $theta=\pi/2$ this gives the usual coplanar 120$^\circ$ order. The ground state manifold parameterized by $(\phi, \theta)$ is $S^1 \times (0,\pi)$. Such a three-sublattice order continues to be the ground state upon turning on $J_{\pm\pm}$: it is easy to check that for any three-sublattice order the ground state energy is independent of $J_{\pm\pm}$.

We performed linear spin wave analysis to study the magnon excitations. First, we find that the spectrum is gapless at $\Gamma$ and K. This gapless branch can be viewed as the goldstone mode of breaking the $S^1$ symmetry: note that although $J_{\pm\pm}$ breaks spin rotational symmetry on the Hamiltonian level, the ground state manifold restores such a symmetry (manifest in the parameter $\theta$); any choice of the ground state breaks this symmetry and leaves a gapless mode at K.

We would like to understand the absence of intensity at K in the inelastic neutron data. We calculate the structure factor $S(\textbf{\textit{q}},\omega)$. Interestingly, the structure factor at K$_1$ can be computed analytically; the three magnon branches and their relative intensities can be obtained in closed form

\begin{equation}
S(\textbf{\textit{q}}_{\textup{K}_1},\omega)=S_0 
\Bigl\{\delta(\omega)g_\perp^2 \sin^4 (\frac{\theta}{2})
+\delta(\omega-E_{\textup{K}_1}^-)g_\perp^2\cos^4(\frac{\theta}{2})
+\delta(\omega-E_{\textup{K}_1}^+) g_z^2 \sin^2 \theta
\Bigr\},
\end{equation}
where $S_0$ is an overall factor. The energies of gapped magnon branches have the explicit expression
\begin{equation}
E_{\textup{K}_1}^\pm = \frac{9}{4}J_\pm (1\pm \cos\theta).
\label{Eq:EK1}
\end{equation}

Equation~\ref{Eq:EK1} provides an explanation for the absence of zero frequency intensity at K$_1$: if the in-plane $g$-factor $g_\perp$ is much smaller than the out-of-plane $g_z=3.66$, and that the order is coplanar with $\theta=\pi/2$, then the intensity ratio of the gapped and the gapless branches satisfy 
\begin{equation}
\gamma_{\textup{K}_1}\equiv \frac{I(\omega=\frac{9}{4}J_\pm)}{I(\omega=0)}=1+4\left(\frac{g_z}{g_\perp}\right)^2 \approx 14.4 \ \textup{for}\ g_\perp =2,
\end{equation}
i.e. the gapped intensity can be one order of magnitude stronger than the gapless one.

One can similarly analyze the intensity ratio of the gapped and the gapless branches at $\Gamma_1$ and $\Gamma_2$, and find that $\gamma=\frac{1}{4}(\frac{g_\perp}{g_z})\approx1/13.4$ for $g_\perp=2$, i.e. for $\Gamma_{1,2}$ the intensity is concentrated at $E$=0.

The simulated structure factors $S(\textbf{\textit{q}}, \omega)$ at $B=0$~T is shown in the Figure~5 in the main text. As can be seen, the structure factor compares well with the experimental inelastic neutron data.

\subsection{\label{appendix: Quantum Effects} Quantum Effects}

The quantum $S=\frac{1}{2}$ model at the parameter point $J_z/J_\pm=-1$ is exactly solvable, which has been briefly discussed in the main text. This model has been studied in detail by Momoi and Suzuki \cite{momoi1992ground}: The quantum ground state manifold exactly coincides with the classical ground state manifold, consisting of (the quantum version of) umbrella orders parameterized by $(\phi,\theta)\in S^1 \times [0,\pi]$. The classical phase boundary $J_z/J_\perp = -1/2,\ J_c = 0$ is also the quantum phase boundary between a FM phase $(|J_z/J_\perp|>1/2)$ and a chiral ordered phase $(|J_z/J_\perp|<1/2)$. The gapless branch of the spin wave has a quadratic dispersion $E \sim k^2$ at K$_1$ point, and the diverging fluctuations of magnetization lead to a non-magnetic ground state. It is shown that neither FM nor chiral order can develop at the point $J_z/J_\perp = -1/2,\ J_c = 0$ at any finite temperatures \cite{momoi1992ground}.

The ground states are two-fold degenerate, with wave functions (in the zero magnetization sector) of the form
\begin{equation}
|\Phi^\pm(\theta,\phi)\rangle = \Bigl(\otimes_{\textbf{\textit{r}}_a} |a,\theta, \phi \rangle_{\textbf{\textit{r}}_a}^\pm
\otimes_{\textbf{\textit{r}}_b} |b,\theta, \phi \rangle_{\textbf{\textit{r}}_b}^\pm
\otimes_{\textbf{\textit{r}}_c} |c,\theta, \phi \rangle_{\textbf{\textit{r}}_c}^\pm\Bigr),
\end{equation}
where for each site $\textbf{\textit{r}}_d$, 
\begin{equation}
|d,\theta,\phi\rangle^\pm \equiv \frac{1}{\sqrt{2}} \left(
|\uparrow\rangle + e^{\pm i \frac{2\pi d}{3}} e^{i\phi} \tan(\frac{\theta}{2})|\downarrow\rangle\right),\quad d=a,b,c=0,1,2.
\end{equation}
Note that the magnetization of the state $|\Phi^\pm(\theta,\phi)\rangle$ depends on the value of $\theta$.

The (elastic) structure factors for the specific state with $\theta=0$ and zero magnetization have been calculated \cite{pal2021colorful}. It is shown that the in-plane component $S^{xx}(\textbf{\textit{q}}) + S^{yy}(\textbf{\textit{q}})$ has intensity at K, while the out-of-plane component $S^{zz}(\textbf{\textit{q}})$ has intensity at $\Gamma$. However, as commented above, the experimental ground state is a disordered one, which presumably is the superposition of $|\Phi^\pm (\theta,\phi)\rangle$ with all values of $\theta$ and $\phi$, and one should not regard the structure factors in Ref.\cite{pal2021colorful} as a direct simulation of the experimental neutron data.


%

